\documentclass[final,5p,times,twocolumn]{elsarticle}

\usepackage{lineno}
\usepackage{makecell}
\usepackage{xcolor}
\usepackage{hyperref}
\usepackage{mathtools}
\usepackage{cleveref}
\crefname{figure}{Fig.}{Figs.}
\crefname{table}{Table}{Tables}
\usepackage{pdfpages}

\definecolor{clearblue}{RGB}{173, 216, 230}

\hypersetup{
    colorlinks=true,
    linkcolor=clearblue,
    urlcolor=black,
}

\usepackage{stfloats}
\usepackage{graphicx}
\usepackage{amsmath}
\usepackage[skip=10pt]{caption}
\begin{document}

\begin{frontmatter}

\title{Organ dose optimization for a point-of-care forearm X-ray photon-counting CT}

\author[1]{Pierre-Antoine Rodesch\corref{cor1}}
\ead{pierreantoine.rodesch@gmail.com}

\author[1]{Anaïs Viry}

\author[1]{Mouad Khorsi}

\author[2]{Fabio Becce}

\author[1,3]{Jérôme Damet\fnref{fn1}}

\author[1]{Lucía Gallego Manzano\fnref{fn1}}

\cortext[cor1]{Corresponding author}
\fntext[fn1]{Both authors contributed equally to this work.}
\fntext[fn2]{Submitted to Physica Medica on 12th May 2025.}

\affiliation[1]{organization={Institut of Radiation Physics (IRA), Lausanne University Hospital (CHUV) and University of Lausanne (UNIL)},
                 city={Lausanne}, 
                 country={Switzerland}}
\affiliation[2]{organization={Department of Diagnostic and Interventional Radiology, Lausanne University Hospital (CHUV) and University of Lausanne (UNIL)},
                city={Lausanne}, 
                country={Switzerland}}
\affiliation[3]{organization={University of Otago},
                city={Christchurch}, 
                country={New Zealand}}

\begin{abstract}
\textit{Background:} Spectral shaping is a computed tomography (CT) dose optimization technique that adjusts source voltage and filtration to reduce patient radiation exposure without compromising image quality. Traditionally, radiation dose has been assessed using the computed tomography dose index (CTDI). However, emerging dosimetric approaches aim to enable patient-specific evaluations by estimating organ absorbed doses, providing a more accurate representation of the biological impact. This study investigates spectral shaping for an extremity photon-counting detector (PCD) CT, through organ absorbed dose estimation and image quality evaluation.

\noindent \textit{Method:} Monte Carlo simulations were conducted to evaluate various combinations of source voltage and filtration. Tube voltage ranged from 80 to 140~kV, combined with three distinct filtration material and thicknesses. Simulations included three stages: a standardized phantom for CTDI assessment, an adult forearm phantom for organ dose measurement, and an image quality phantom for evaluation of an advanced image quality metric: the detectability index.

\noindent \textit{Results:} In a wrist PCD-CT imaging protocol, operating the source at 80~kV can reduce the radiation dose by up to 50\%. This reduction is achieved while maintaining the same detectability index value as the standard 120~kV protocol. However, the optimal filtration depends on the organ targeted for dose reduction, as bone and skin benefit from opposing filtration approaches. While CTDI provides a useful initial estimate, it may lead to suboptimal optimization compared to organ-specific dose evaluation.

\noindent \textit{Conclusions:} Patient-specific dosimetry based on organ absorbed dose estimation offers a more accurate framework for optimizing CT protocols through spectral shaping than conventional CTDI-based approaches.
\end{abstract}


\end{frontmatter}

\section{Introduction}
\label{sec:intro}

Computed tomography (CT) is an imaging technique that utilizes X-rays to reconstruct a three-dimensional volume of the scanned object. The number of CT examinations performed annually is continually increasing. Despite accounting for less than 10\% of all examinations, CT contributes to more than half of the medical exposure dose~\cite{UNCSEAR_2020_2021,smith2025projected}. The International Commission on Radiological Protection (ICRP) classifies the cancer risk from CT radiation as ranging from "very low" to "low", depending on the specific CT protocol used~\cite{harrisonUseDoseQuantities2021}. These radiation dose levels can be an order of magnitude higher than those from conventional radiography procedures~\cite{harrisonUseDoseQuantities2021}. From an epidemiological perspective, there is growing evidence linking cancer risk to irradiation doses, even below 100~mGy~\cite{ruhmCancerRiskFollowing2022, laurierScientificBasisUse2023}. Further research is needed to better understand the impact of low-dose radiation~\cite{tapioIonizingRadiationinducedCirculatory2021}. Moreover, recurring CT examinations can result to high cumulative dose levels and an increased associated cancer risks~\cite{brambillaMultinationalDataCumulative2020,frijaCumulativeEffectiveDose2021}. Across all cumulative dose levels, patient survival rates are often overlooked, highlighting the need to strengthen patient radioprotection~\cite{mataacWhatProportionCT2024}.

During a CT examination, the dose is reported using the CT dose index volume (CTDI\textsubscript{vol}), measured with a standardized poly-methyl-methacrylate (PMMA) phantom~\cite{anderssonEstimatingPatientOrgan2019}. This metric is useful for comparing different CT devices but has limitations: it is not fully correlated with patient dose and is not patient-specific~\cite{mccolloughCTDoseIndex2011}. In CT imaging, a shift from scanner beam dosimetry to patient-specific dose estimates is foreseen, emphasizing the need for more patient-specific dose indicators~\cite{bottollier-depoisVisionsRadiationDosimetry,anderssonEstimatingPatientOrgan2019}. This involves tracking radiation dose for each patient individually~\cite{brambillaMultinationalDataCumulative2020}, rather than using population-averaged numerical phantoms such as the ICRP reference computational phantoms~\cite{icrp2009adult,kimAdultMeshtypeReference2020}. The most accurate approach to achieving patient-specific dosimetry is through the evaluation of organ absorbed dose~\cite{rehaniDoseDoseDose2024, damilakisCTDosimetryWhat2021}. organ absorbed dose is the recommended metrics to quantify the patient dose~\cite{ anderssonEstimatingPatientOrgan2019} and estimate the risk to health from radiation exposure~\cite{harrisonEffectiveDosesRisks2023}. Organs exhibit different dose-related risks, with risk levels depending on patient age and gender~\cite{harrisonEffectiveDosesRisks2023}. In CT examinations, organ dose also varies, with deeper organs generally absorbing lower doses compared to superficial ones~\cite{yamashitaDirectMeasurementRadiation2021}. Monte Carlo (MC) simulations are the gold-standard technique for computing organ absorbed dose~\cite{sechopoulosMonteCarloReference2015, sechopoulosRECORDSImprovedReporting2018, akhavanallafAssessmentUncertaintiesAssociated2020}. This simulation tool generates a dose distribution map~\cite{somasundaramDevelopmentValidationOpen2019}, from which the organ absorbed dose can be computed.

\begin{figure*}[hb]
\centering
\includegraphics[width=.9\textwidth]{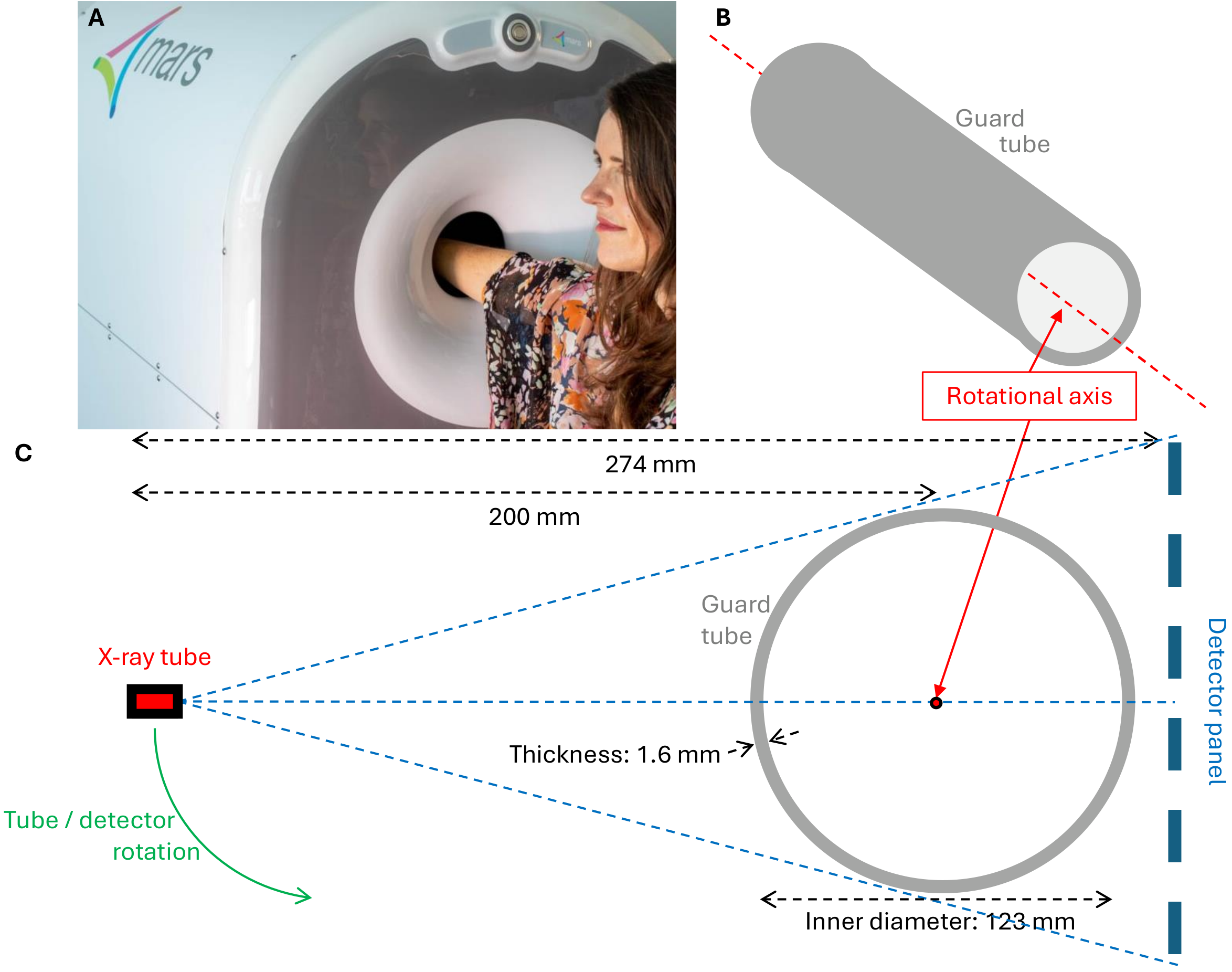}
\caption{A: Illustrative picture of the MARS Extremity 5X120 PCD-CT scanner. B: 3D representation of the guard tube. C: Detailed geometry implemented in the MC simulation.}
\label{fig:mars_geometry}
\end{figure*}

The dose distribution map depends on the X-ray source parameters, such as the X-ray tube filtration and voltage. Modifying these parameters to optimize the CT protocol is defined as spectral shaping~\cite{wong2022spectral} and has been explored with new filtration materials, such as tin~\cite{greffierBrainImageQuality2023,hasegawa2022tin,grunz2022spectral}. The benefits of spectral shaping are linked to improvements in the latest image denoising techniques, such as iterative reconstruction or AI-based denoising algorithms~\cite{jiangDeepLearningReconstruction2022, racineTaskbasedCharacterizationDeep2020, muckSaferImagingComparative2024}. Indeed, spectral shaping results in different noise properties that can be counterbalanced by an effective noise reduction technique. Previous studies~\cite{Reza_Ay_2013,steidelDoseReductionPotential2022,greffierBrainImageQuality2023} have demonstrated that specific spectral shaping could improve image quality at constant CDTI\textsubscript{vol} but they have not investigated organ absorbed dose. It has been demonstrated that optimizing the CTDI\textsubscript{vol} value does not necessarily imply the optimization of organ absorbed dose. Fong \textit{et al.}~\cite{fongEffectTinSpectral2024} have demonstrated that tin filtration can reduce the CTDI\textsubscript{vol}. But the dose conversion coefficient (DCC), which converts the CTDI\textsubscript{vol} into absorbed dose, increases with the CTDI\textsubscript{vol} for some organs, mitigating the dose reduction. However, no imaging quality metrics were studied in their work.

Photon-counting detectors (PCDs) represent an emerging technology that is revolutionizing CT imaging. They offer superior spatial resolution, higher contrast and reduced noise levels compared to conventional energy-integrating detector (EID) CT scanners~\cite{si_mohamed_PCCT,Rajendran_2021,Zhan_2023}. When combined with spectral shaping and advanced noise reduction techniques, PCD-CT unlocks efficient low-dose protocols, making them highly suitable for large-scale screening programs. In this context, optimizing dose is essential. Moreover, a PCD-CT provides a higher contrast than an EID-CT because of the difference in detection technique~\cite{si2021spectral}. This could potentially change the spectral shaping optimal parameters. Leveraging PCDs, MARS Bioimaging Ltd. (Christchurch, New Zealand) has developed a upper extremity scanner featuring a small field-of-view (120~mm) (\hyperref[fig:mars_geometry]{\cref{fig:mars_geometry}}) and a high spatial resolution~\cite{gallego2023clinical}. The primary application of the scanner is a wrist protocol~\cite{Mars_White_Paper}, for either the diagnosis of crystal arthropathies such as gout~\cite{huber2021differentiation,stamp_2019} or the detection of small bone fractures~\cite{mourad2024chances}. The scanner is designed for point-of-care utilization and specific clinical applications requiring ultra high resolution, at the cost of a potential longer acquisition time (a few minutes). Given its novelty and specificity, analyzing the dose is significant and insightful.    

The objective of this work was to find the optimal spectral shaping (source voltage and filtration combination) to reduce the dose for a dedicated forearm PCD-CT device. In this study, the dose is not evaluated using the CTDI\textsubscript{vol} but directly the organ absorbed dose. A wrist protocol for different source parameters was simulated with the FLUKA MC code as distributed by the FLUKA CERN collaboration~\cite{AhdidaFlukaCern2022,battistoni2015overview,hugo2024latest}. For the wrist, the organs have been classified into three categories: bone, soft tissue, and skin. The simulation evaluates the organ absorbed dose and the corresponding image quality, assessed through the detectability index. To our knowledge, this is the first study to link organ dose with a comprehensive image quality metric. From the simulation results, it can be determined which source parameters provide the smallest organ dose for a given image quality level. Additionally, the CTDI\textsubscript{vol} was also evaluated to investigate how its optimization is related to organ absorbed dose reduction.

The first section of this article describes the MC simulations developed for this study for CTDI\textsubscript{vol}, organ absorbed dose and image quality evaluation. The second section presents the results that are discussed in the last section. 

\section{Material and Methods}

\subsection{Scanner geometry}
\label{sec:scanner_geometry}

\begin{figure*}[hb]
\centering
\includegraphics[width=\textwidth]{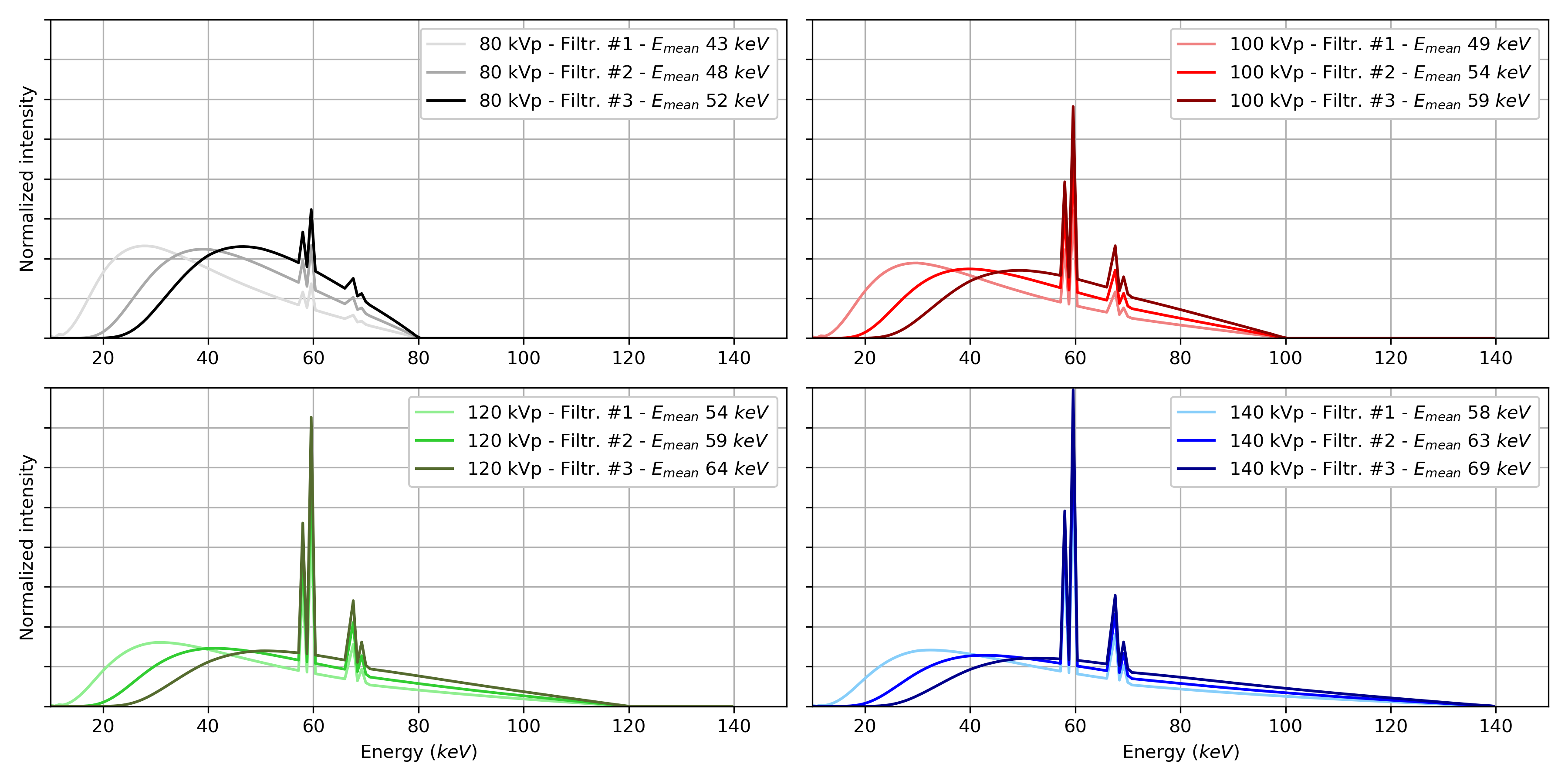}
\caption{The incident source spectra investigated in this work for 1~mAs. The same figure representing the spectra normalized as a density is provided in supplemental figure~1.}
\label{fig:spectra}
\end{figure*}

The MC simulation replicates the geometry of the MARS Extremity 5X120 (MARS Bioimaging Ltd., Christchurch, New Zealand)~\cite{gallego2023clinical}. In this configuration, the forearm is held in place by a guard tube (\hyperref[fig:mars_geometry]{\cref{fig:mars_geometry}.A}). The guard tube is a 1.6~mm thick aluminum cylinder with an inner diameter of 123~mm (\hyperref[fig:mars_geometry]{\cref{fig:mars_geometry}.B} and \hyperref[fig:mars_geometry]{C}). This guard tube was included in all simulations. An X-ray source and a detector panel rotate 360 degrees around the guard tube to acquire projection images. The source-to-isocenter and source-to-detector distances are 200~mm and 276~mm, respectively. In the simulation, the detector was modeled as a flat panel with 1792$\times$128 pixels, a pixel pitch of 0.11~mm, and a pixel thickness of 0.3~mm. Due to the relatively small size of the scanned objects, a distinctive feature of the scanner is the absence of an anti-scatter grid (ASG).

\subsection{Spectral shaping}

\begin{figure*}[t]
\centering
\includegraphics[width=\textwidth]{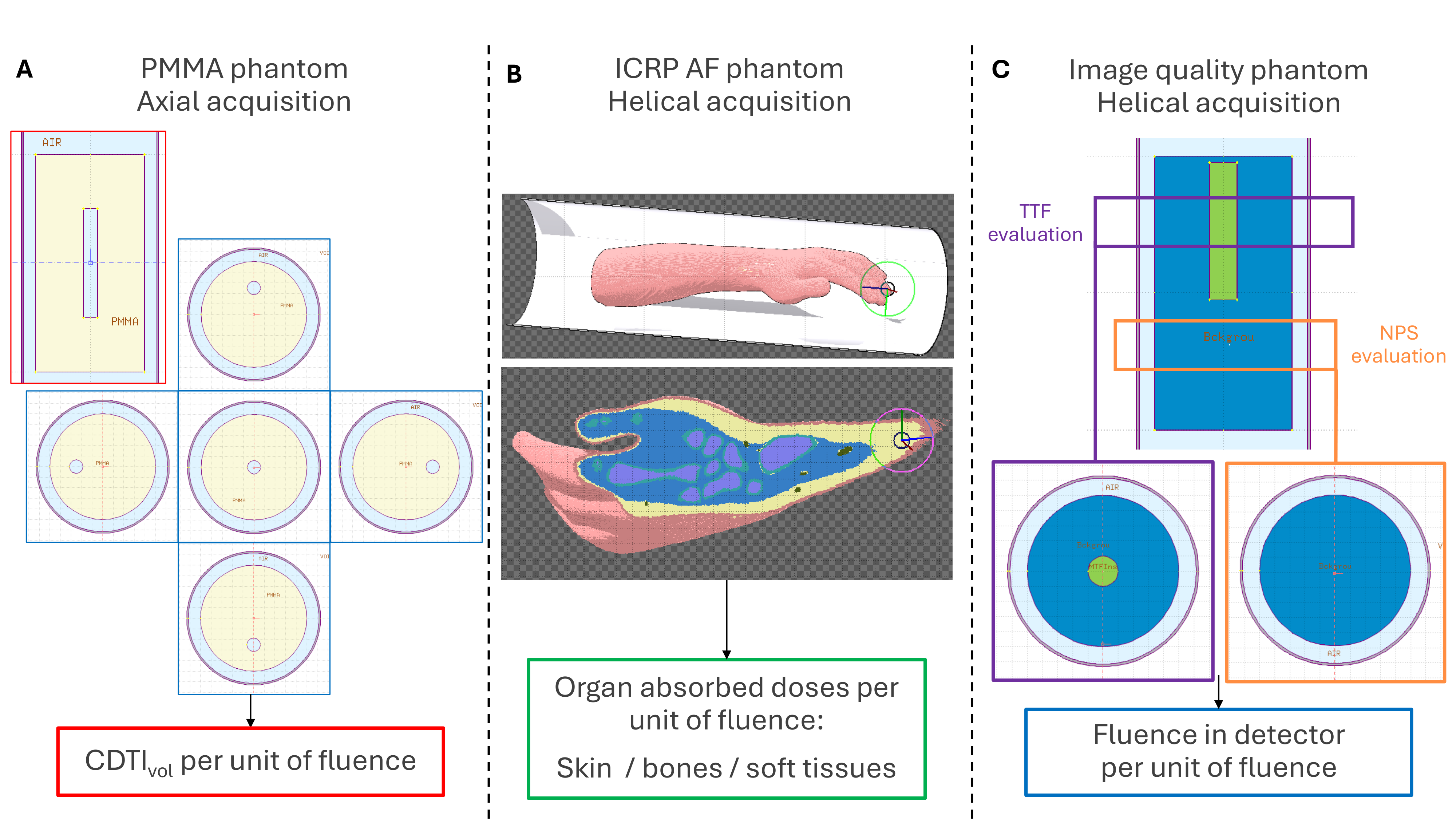}
\caption{A: Phantom scheme used for the CTDI\textsubscript{vol} simulation. B: Illustration of the ICRP forearm phantom~\cite{kimAdultMeshtypeReference2020} (AF: adult female). C: Phantom scheme used for the image quality evaluation. These images were generated with the help of the FLAIR3 user interface~\cite{donadon2024flair3}.}
\label{fig:simulation}
\end{figure*}

One of the inputs to the MC simulation is the X-ray source spectrum. In this study, four source voltages (80, 100, 120, and 140~kV) were combined with three different filter configurations, mixing aluminum and copper filters, as indicated in \hyperref[tab:spectra]{\cref{tab:spectra}}. The spectra were generated using the SpekPy simulation tool~\cite{poludniowskiTechnicalNoteSpekPy2021}. These filter configurations represent additional filtration and do not include the 1.6~mm aluminum filtration introduced by the guard tube. The filters were selected to achieve equivalent aluminum half-value layers (HVL, expressed in mmAl) of 4, 6, and 8~mm at 120~kV at the isocenter, including attenuation from the guard tube. The resulting spectra, scaled to 1~mAs at a distance of 20~cm from the source, are shown in \hyperref[fig:spectra]{\cref{fig:spectra}}. However, in the FLUKA CERN code, the input spectra are normalized as photon density, as shown in supplemental figure~1. Spectral properties are presented in \hyperref[tab:spectra]{\cref{tab:spectra}}. In \hyperref[tab:spectra]{\cref{tab:spectra}}, the HVL and mean energy of the spectra were calculated including the inherent 1.6mm aluminum filtration from the guard tube. The spectrum of the scanner for the routine protocol corresponds to the 120~kV spectrum with a 6~mmAl HVL~\cite{gallego2023clinical}, defined as 120kV-Filtr~\#2.

\begin{table}[h]
\centering
\caption{Incident spectra properties. mmAl: equivalent aluminum millimeters. \\ *these values were computed with an extra 1.6 mm aluminum filtration corresponding to the guard tube wall.}
\setlength{\tabcolsep}{5pt}
\resizebox{0.98\columnwidth}{!}{
\begin{tabular}{|l|c|c|c|c|}
\hline
Name & \makecell{Voltage \\ (kVp)} & \makecell{Filtration \\ (mm)} & \makecell{$E_{mean}$* \\ (keV)} & \makecell{HVL* \\ (mmAl)} \tabularnewline
\hline
80kV-Filtr\#1 & 80 & Cu: 0.0 - Al: 1.0 & 42.7 & 2.6 \tabularnewline
80kV-Filtr\#2 & 80 & Cu: 0.1 - Al: 1.0 & 47.8 & 4.3 \tabularnewline
80kV-Filtr\#3 & 80 & Cu: 0.2 - Al: 3.4 & 51.7 & 5.7 \tabularnewline
\hline
100kV-Filtr\#1 & 100 & Cu: 0.0 - Al: 1.0 & 48.8 & 3.3 \tabularnewline
100kV-Filtr\#2 & 100 & Cu: 0.1 - Al: 1.0 & 53.6 & 5.1 \tabularnewline
100kV-Filtr\#3 & 100 & Cu: 0.2 - Al: 3.4 & 58.5 & 7.0 \tabularnewline
\hline
120kV-Filtr\#1 & 120 & Cu: 0.0 - Al: 1.0 & 53.9 & 4.0 \tabularnewline
120kV-Filtr\#2 & 120 & Cu: 0.1 - Al: 1.0 & 58.8 & 6.0 \tabularnewline
120kV-Filtr\#3 & 120 & Cu: 0.2 - Al: 3.4 & 63.9 & 8.0 \tabularnewline
\hline
140kV-Filtr\#1 & 140 & Cu: 0.0 - Al: 1.0 & 58.3 & 4.7 \tabularnewline
140kV-Filtr\#2 & 140 & Cu: 0.1 - Al: 1.0 & 63.4 & 6.8 \tabularnewline
140kV-Filtr\#3 & 140 & Cu: 0.2 - Al: 3.4 & 68.7 & 8.8 \tabularnewline
\hline
\end{tabular}}
\label{tab:spectra}
\end{table}

The spectra exhibit different mean energy values. As a polychromatic photon beam is attenuated through an object, its mean energy increases and this phenomenon is known as beam-hardening (BH). One of the impacts of the mean energy is on image contrast, as well as on the associated BH. A lower mean energy generally results in higher contrast and, consequently, more pronounced BH that can lead to artifacts in the presence of highly attenuating structures~\cite{ketcham2014beam,park2015computed}. The contrast will be evaluated in the CT images in the following sections. The theoretical BH effect through 15~cm of water (corresponding to the water equivalent diameter (WED) of a forearm) was also evaluated by computing a BH factor. It was defined as the difference in water linear attenuation coefficients (LAC). The LAC was evaluated at the mean energy of the incident spectrum and at the mean energy of the same spectrum after attenuation through 15~cm of water. This difference was then converted into Hounsfield Units (HU). No BH correction was implemented in this work before the CT reconstruction.

\subsection{MC simulation framework}

Different codes have been developed to perform MC simulations. In this work, the simulations were performed using the FLUKA CERN code (version 4-4.1)~\cite{AhdidaFlukaCern2022,battistoni2015overview,hugo2024latest}, which has previously been applied to CT dose evaluation and positively benchmarked~\cite{battistoni2016fluka,somasundaramDevelopmentValidationOpen2019} against the reference datasets provided by the American Association of Physicists in Medicine (AAPM) Task Group 195~\cite{sechopoulosMonteCarloReference2015}. All computations were performed with the FLUKA default parameters (FLUKA "DEFAULTS" card set to "PRECISIOn"). The photon transport threshold was configured to 10~keV and the electron transport to the maximum incident energy, as electrons travel a negligible distance and
deposit their energy in the same voxel as the ionization process~\cite{podgorsak2003review,wangFastLinearBoltzmann2019,NIST_database}. All simulations were executed on a high-performance computing cluster, enabling a statistical uncertainty of less than 5\%. The simulations were divided into three cases (\hyperref[fig:simulation]{\cref{fig:simulation}}): the first corresponds to the CTDI\textsubscript{vol} computation, the second to the absorbed dose evaluation, and the third to the image quality assessment. All three cases were simulated for the 12 spectra presented in \hyperref[fig:spectra]{\cref{fig:spectra}}. The input X-ray source was modeled using an energy spectrum and a 0.2~mm focal spot size. It was collimated to match detector panel size.

\subsection{CTDI\textsubscript{vol}}
Traditionally, the CTDI\textsubscript{vol} is defined using either a 16~cm (head phantom) or 32~cm (body phantom) diameter PMMA cylinder. For the current geometry, it is adapted to a 10~cm diameter phantom~\cite{gallego2023clinical}, as shown in \hyperref[fig:simulation]{\cref{fig:simulation}.A}. An axial protocol with 360 projections was simulated. The fluence was recorded within a 10~mm diameter, 10~cm height cylinder at five positions: the center and the four peripheral positions. It was then converted into air kerma using the conversion coefficients provided in the ICRP Publication~74~\cite{international1996icrp}. A weighted CTDI\textsubscript{w} was computed from the central and peripheral values. The final CTDI\textsubscript{vol} value was obtained by dividing CTDI\textsubscript{w} by the pitch. In the helical acquisitions for the wrist protocol, the pitch was set to 0.58.

The simulation was run with $2 \times 10^{9}$~primaries over 5~cycles. The output of this first simulation was a CTDI\textsubscript{w} value per primary for each spectrum. From this value, the fluence corresponding to CTDI\textsubscript{w} levels of 1, 2, 5, 10, 20, 50, and 100~mGy was computed. These levels are then used for the image quality simulations and are defined as $P^c_s$, where $c$ denotes the CTDI\textsubscript{w} level and $s$ the spectrum index. In this work, $P^c_s$ is a number of primaries per projection to reach the CTDI\textsubscript{w} level $c$.

\subsection{Organ absorbed dose}
\begin{figure}[h]
\centering
\includegraphics[width=\columnwidth]{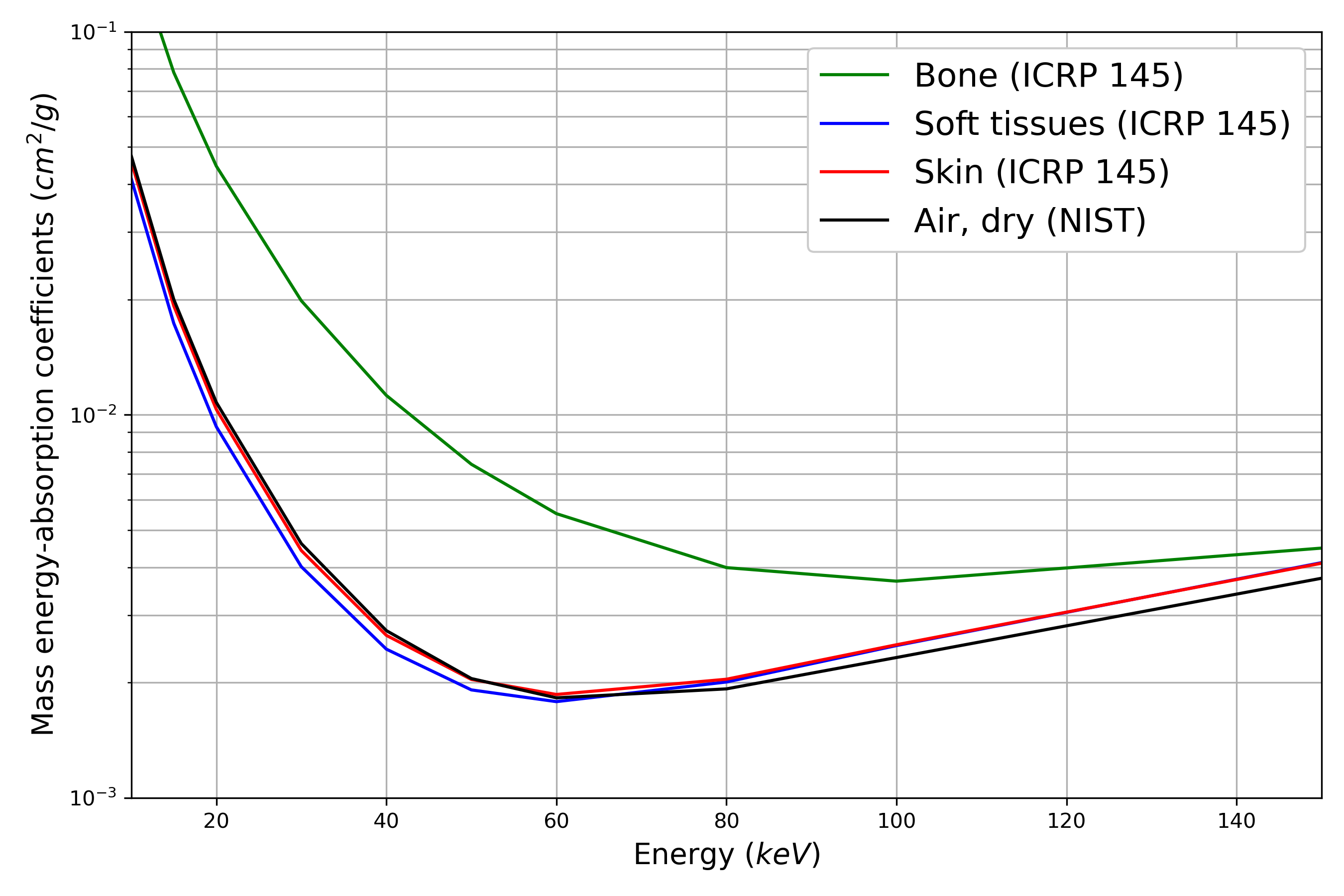}
\caption{Bone, soft tissue, skin and air mass energy absorption coefficients. (For interpretation of the references to colour in this figure legend, the reader is referred to the web version of this article.)}
\label{fig:ener_absorb}
\end{figure}

The organ absorbed dose was evaluated by simulating the MARS Extremity 5X120 wrist protocol~\cite{gallego2023clinical} using the ICRP adult female mesh phantom~\cite{kimAdultMeshtypeReference2020}. The wrist protocol is a helical acquisition with a pitch of 0.58 and 360~projections per rotation. The acquisition length was set to 4~cm to image the entire wrist joint as well as the distal portions of the ulnae and radii, resulting in seven rotations of the X-ray source around the phantom.  
At the moment of these simulations, the FLUKA CERN code did not yet support mesh phantoms directly. Therefore, the ICRP mesh phantom was converted into a voxel phantom with an isotropic voxel size of 0.4~mm~\cite{dong2023simple}. The aluminum guard tube was added to the voxelised volume (\hyperref[fig:simulation]{\cref{fig:simulation}.B}). The final phantom contains 11~structures, each assigned a material defined in ICRP Publication 145~\cite{kimAdultMeshtypeReference2020}. For organ dose estimation, three categories were retained: bone, soft tissue, and skin (\hyperref[tab:material]{\cref{tab:material}}). The corresponding averaged mass absorption coefficient for each category are displayed in (\hyperref[fig:ener_absorb]{\cref{fig:ener_absorb}})
A dose distribution volume was scored as the output of the MC simulation, using the same voxel geometry as the voxelised phantom. The simulation was run with $1\times10^9$ primaries over 5~cycles. Organ doses were then computed by applying tissue-specific masks derived from the voxelised phantom. An example of dose map distribution and corresponding organ mask are provided in supplemental figure~3. 

\begin{table}[!h]
\centering
\caption{List of forearm phantom structures and associated materials. The first and second columns correspond to the structure and material names from ICRP Publication 145~\cite{kimAdultMeshtypeReference2020}.\\
\textit{*The medullary cavity was included in the bone scoring category but is not visible in the wrist protocol.}
}
\setlength{\tabcolsep}{5pt}
\resizebox{0.98\columnwidth}{!}{
\begin{tabular}{|l|c|c|c|c|}
\hline
ICRP Organ/Tissue & Medium & Category \tabularnewline
\hline
Background & Air &  \tabularnewline
Blood in large arteries, arms & Blood &  \tabularnewline
Blood in large veins, arms & Blood &  \tabularnewline
Ulnae and radii, cortical & Cortical bone & Bone \tabularnewline
Ulnae and radii, spongiosa & Spongiosa & Bone \tabularnewline
Ulnae and radii, medullary cavity & Medullary cavity & Bone* \tabularnewline
Wrists and hand bones, cortical & Cortical bone & Bone \tabularnewline
Wrists and hand bones, spongiosa & Spongiosa & Bone \tabularnewline
Muscle & Muscle & Soft tissue \tabularnewline
RST & RST & Soft tissue \tabularnewline
Skin & Skin & Skin \tabularnewline
Guard tube & Aluminum &  \tabularnewline
\hline
\end{tabular}}
\label{tab:material}
\end{table}

\subsection{Image quality}
The image quality was asserted through the MC simulation of an image quality phantom (\hyperref[fig:simulation]{\cref{fig:simulation}.C}). Even though image quality is mainly impacted by geometrical factors, it also depends on the scatter magnitude. The studied scanner is not equipped with an ASG, and the scatter magnitude depends on the incident spectrum. That is why the X-ray projections were computed with a MC code to take into account the scattered X-rays. The image quality phantom is a 100~mm diameter cylinder containing a 20~mm diameter insert with the material spongiosa. The background material was set to water. The same helical protocol as for the ICRP phantom was simulated. In this simulation, the detector panel was inserted in the geometry. The photon fluence was scored in each pixel of the detector panel and defined as $N_s$, where $s$ is the incident spectrum index.

An additional projection simulation was computed to estimate the photon flux without the phantom to have a raw flux reference, defined as $N^0_s$. The output of the MC simulation were a fluence per primary for each spectrum case. As the pixel area is small ($0.11 \times 0.11$~mm), this simulation required a high computation time, and each projection was run with $2.10^{9}$ primaries and 5~cycles, leading to a long computation time when multiplied by the number of projections.

These fluences were scaled by the result of the CTDI$_{w}$ simulation to obtain a number of photons corresponding to a selected CTDI$_{w}$ level $c$:
\begin{equation}
	F^c_s = P^c_s \times N_s
\end{equation}
\begin{equation}
	F^{0c}_s = P^c_s \times N^0_s 
\end{equation}
A poissonnian noise was added to $F^c_s$ and not the 
raw flux. The attenuation was then computed as:
\begin{equation}
	p^{c}_s = -log \frac{F^{c}_s}{F^{0c}_s}
\end{equation}
An additional noise free attenuation was calculated for spatial resolution evaluation.

The attenuation was reconstructed with an iterative reconstruction (IR) and a conjugate gradient algorithm, implemented within the Reconstruction Toolkit (RTK)~\cite{rit2014reconstruction,rit2024reconstruction}. The objective function was defined as least-square formulation with a spatial regularization penalization:
\begin{equation}
	L(\mu) = ||p-A\mu||^2_2 + \gamma ||\nabla \mu||^2_2
\end{equation}
where $L$ is the objective function, $\mu$ the LAC to reconstruct, $\lambda$ is the regularization parameter and $\nabla$ the discrete gradient operator.  The second part is called the Tikhonov regularisation and its level is decided by the value of $\lambda$; the greater the lambda, the greater the denoising of the reconstructed image. The $\lambda$ value was set to 10, offering a trade off between noise reduction and spatial resolution preservation. An example of a CT reconstructed volume is presented in supplemental figure~4 for a CTDI\textsubscript{vol}=2~mGy.

The LAC images were reconstructed in $\mathrm{mm}^{-1}$ and then converted to HU using the mean water background value. Different metrics were evaluated on the images: the contrast, the noise power spectrum (NPS), the task transfer function (TTF) and finally, the detectability index $d´$. The latter is a positive number where $d'=0$ represents an impossible detection and infinity represents a perfect detection~\cite{burgess1994statistically,ott2014update,solomon2016correlation}. The contrast and the TTF were measured on the noise-free reconstruction. For each metric, the final value was averaged over 25~slices.

The contrast was defined as the difference between the mean value in the spongiosa bone insert and the mean value in the water background of the phantom.

The NPS and TTF curves were measured using an in-house software~\cite{monnin2020slice}. The NPS was measured from a 51.2~mm square region of interest (ROI), centered in the homogeneous part of the phantom (NPS area in \hyperref[fig:simulation]{\cref{fig:simulation}.C}). The TTF was computed in all 360~degree directions from the spongiosa bone/water edge. The final metric used to estimate image quality was the detectability index $d´$ for a 0.25~mm diameter two-dimensional cylindrical object. It was computed using the non-prewhitening (NPWE) model observer~\cite{burgess1994statistically,ott2014update,solomon2016correlation}:
\begin{equation}
	d' = \frac{\sqrt{2\pi}\;|\Delta HU| \int\limits_{f=0}^{f_{Ny}}S^2(f){TTF}^2(f){VTF}^2(f)fdf}{\sqrt{\displaystyle\int\limits_{f=0}^{f_{Ny}}S^2(f){TTF}^2(f)NPS(f){VTF}^4(f)fdf}} 
\end{equation}
where $|\Delta HU|$ is the contrast, $f$ is the frequency variable, $f_{Ny}$ the radial Nyquist frequency, $S(f)$ is the spectrum object and $VTF(f)$ is the visual transfer function of the human eye, which parameters were selected according to Eckstein~\textit{et al.}~\cite{Eckstein2003}. The detectability index was evaluated for the 12~source parameters sets and the 7~CTDI\textsubscript{vol} levels. From these values were interpolated  for each incident spectrum the CTDI\textsubscript{vol} and organ dose values corresponding to a $d' = 2$. This translates into an area under the receiver operating characteristic (ROC) curve of 0.9~\cite{barrett2013foundations}. This area under the curve (AUC) value indicates a high image quality score. As the score obtained on phantom images is usually higher than on patient images, it is preferable to set a high targeted value for the detectability index.

\vspace{1cm}

\section{Results}
\label{sec:results}

\subsection{CTDI\textsubscript{vol}}

\begin{figure}[h!]
\centering
\includegraphics[width=\columnwidth]{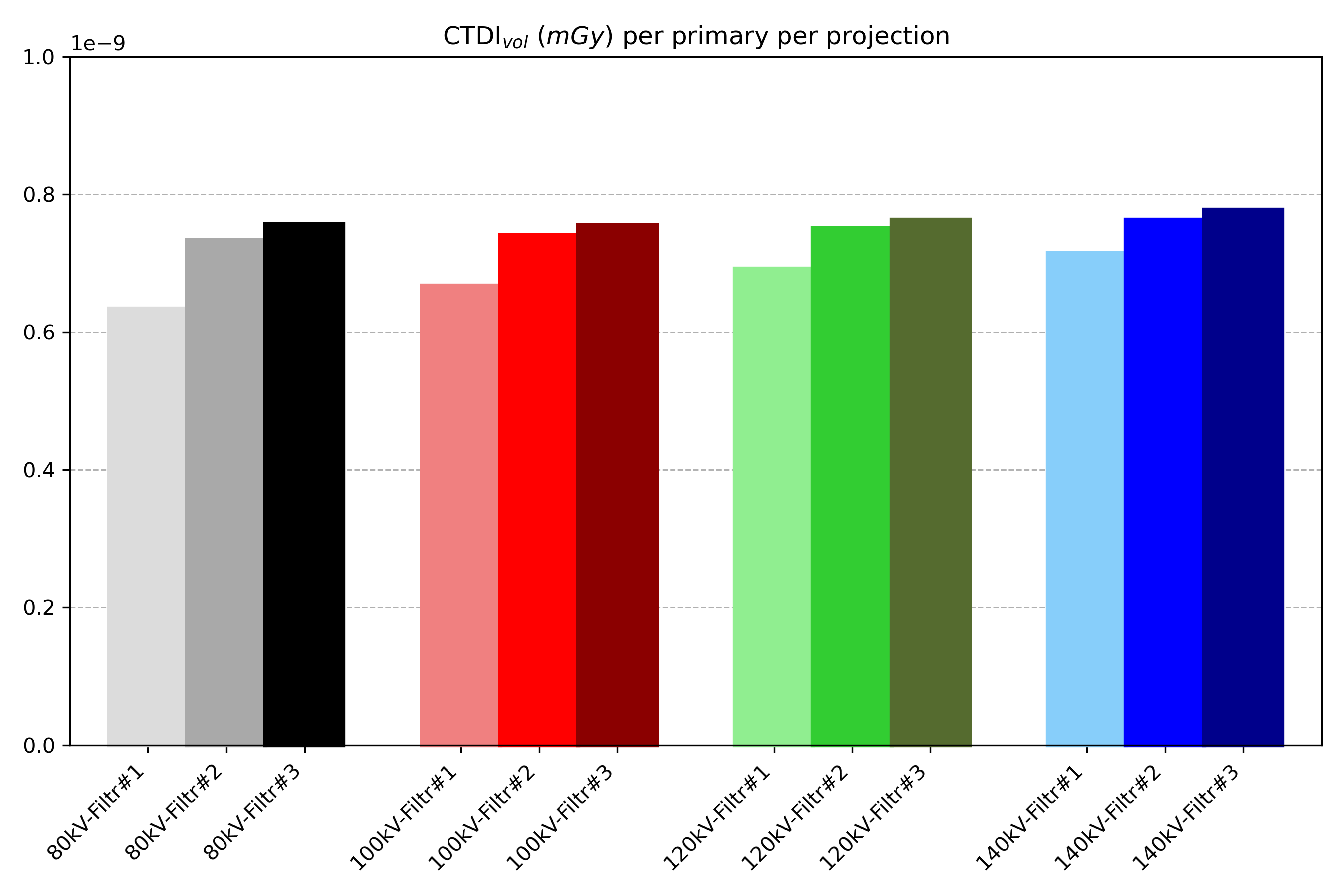}
\caption{Output of the PMMA phantom MC simulation: normalized CTDI\textsubscript{vol} value per primary per projection for all incident spectrum cases.}
\label{fig:ctdi_pre_primary}
\end{figure}

The results of the CTDI\textsubscript{vol} simulation are displayed in \hyperref[fig:ctdi_pre_primary]{\cref{fig:ctdi_pre_primary}} For each spectrum, the CTDI\textsubscript{vol} per primary per projection varies slightly. Alternatively, this can be interpreted as the number of primaries required to achieve 1~mGy also differing. It is important to note that this number is particularly relevant in the context of MC simulations. However, in an experimental setting, the exposure value (expressed in mAs) required to reach 1~mGy would be of primary interest; this value was simulated and is provided in the supplemental figure~2. The values shown in \hyperref[fig:ctdi_pre_primary]{\cref{fig:ctdi_pre_primary}} are then used to determine the number of primaries per projection needed to reach CTDI\textsubscript{vol} levels of 1, 2, 5, 10, 20, 50, and 100~mGy. These values enable the computation of the organ absorbed dose at each CTDI\textsubscript{vol} level, and the corresponding fluence at the detector for the image quality analysis.

\vspace{1cm}

\subsection{Organ absorbed dose}
\begin{figure}[h!]
\centering
\includegraphics[width=\columnwidth]{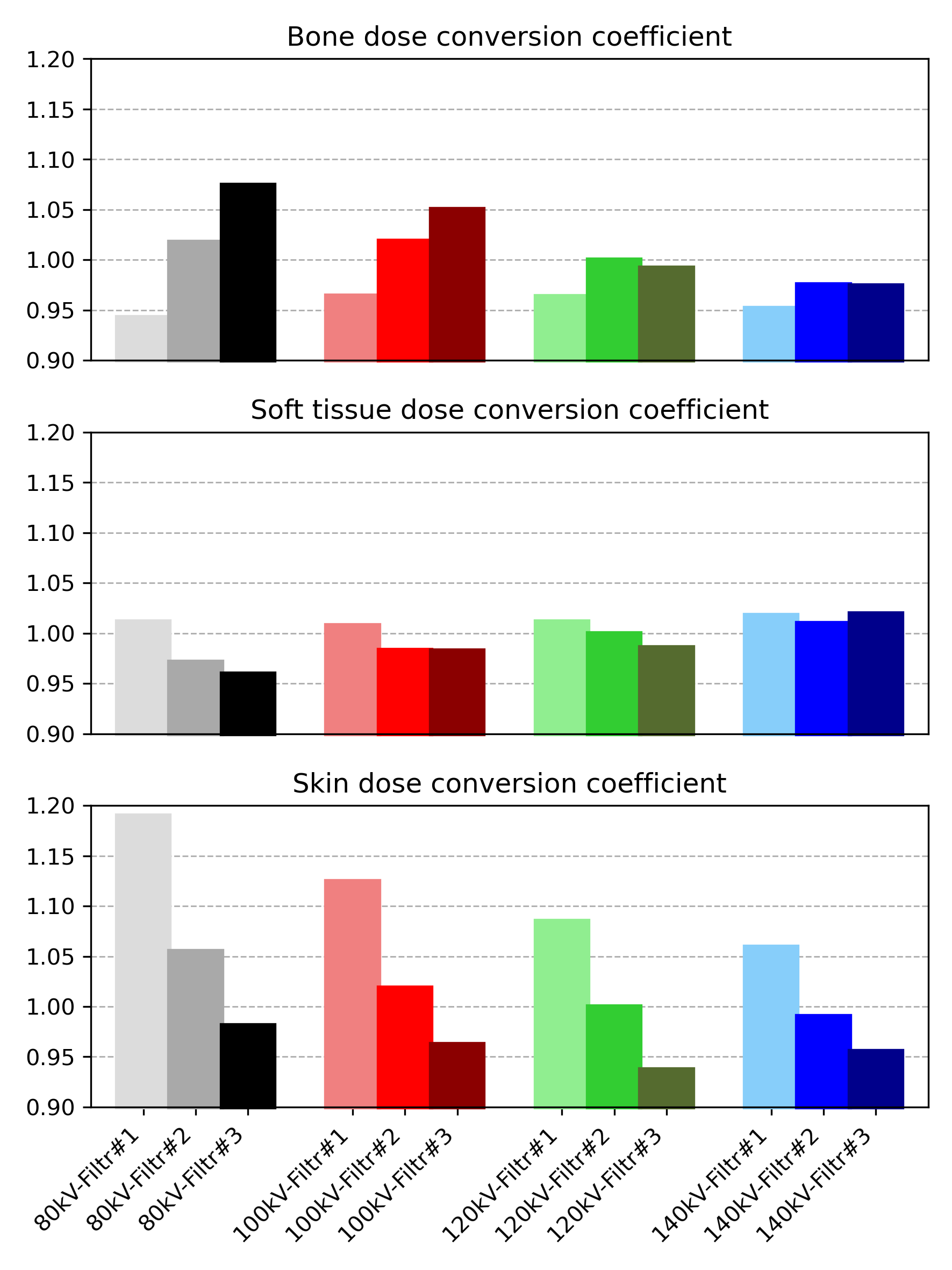}
\caption{Bone, soft tissue and skin absorbed dose per CTDI\textsubscript{vol}, normalized to the 120~kV-Filtr\#2 case.}
\label{fig:organ_dose}
\end{figure}

The organ absorbed dose results are represented using the DCC, defined as the organ absorbed dose per unit of CTDI\textsubscript{vol}. The results were normalized to the standard wrist protocol (120~kV-Filtr\#2). The DCC values for bone, soft tissue, and skin are displayed in \hyperref[fig:organ_dose]{\cref{fig:organ_dose}}. This demonstrates that each material exhibits distinct behavior. The soft tissue DCC remains within 5\% of the standard protocol value across all spectra. In contrast, the bone and skin DCCs show greater variability and opposite trends. Bone benefits from minimal filtration and is negatively affected by high filtration at low voltages, displaying its maximum DCC at 80~kV with Filtration~\#3. In contrast, the skin consistently benefits from increased filtration across all voltages.

\subsection{Image quality}

\subsubsection{Contrast and beam-hardening}
\begin{figure}[h]
\centering
\includegraphics[width=\columnwidth]{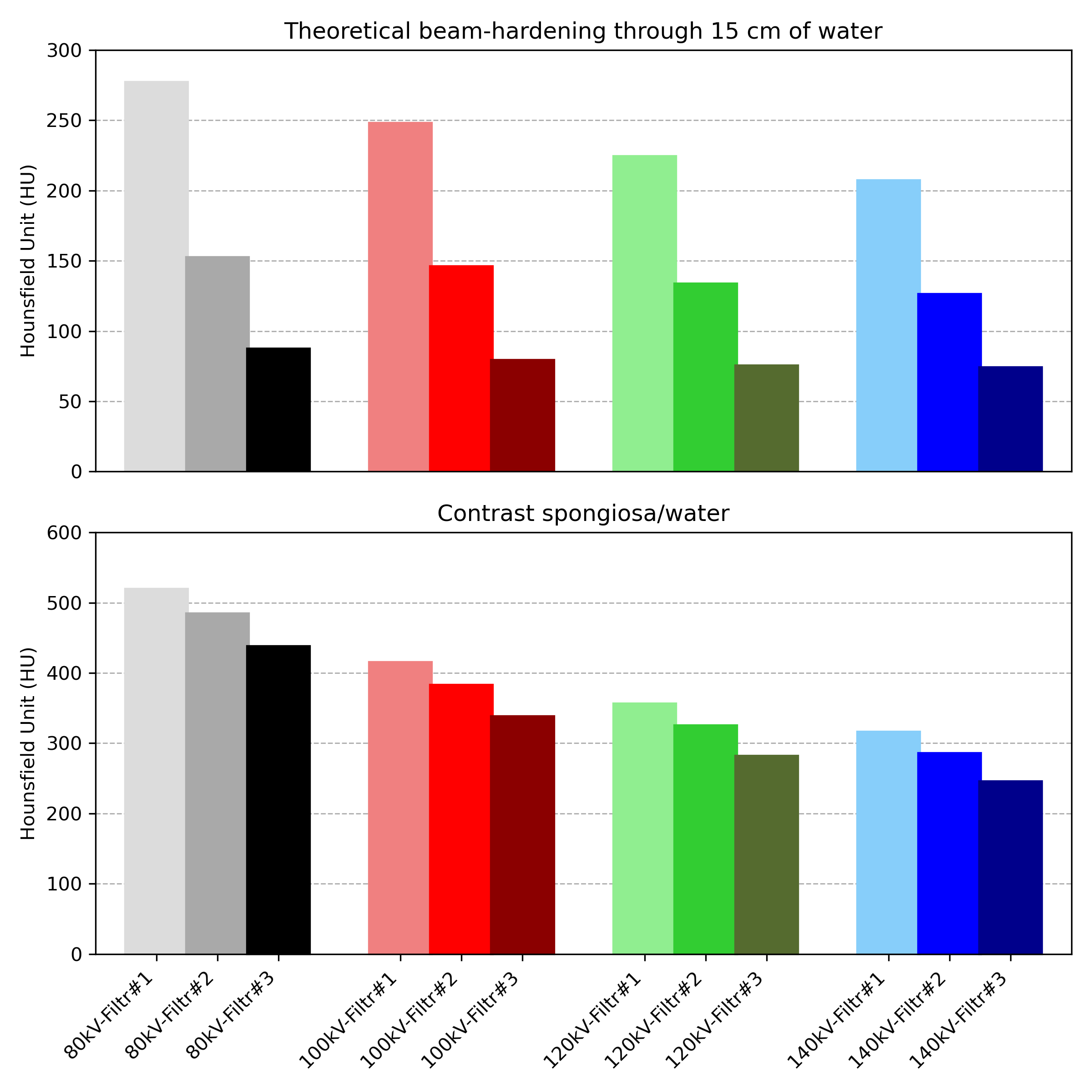}
\caption{Top: theoretical BH values. Bottom: measured contrast in the TTF area of the image quality phantom.}
\label{fig:contrast_bh}
\end{figure}

The contrast and theoretical BH factor through 15~cm of water are presented in \hyperref[fig:contrast_bh]{\cref{fig:contrast_bh}}. These values are directly correlated with the spectrum mean energy presented in \hyperref[tab:spectra]{\cref{tab:spectra}}. For the lowest source filtration, the BH value can reach up to 300~HU and increases more steeply than the contrast. At 80~kV, the contrast is between 1.4 and 1.8 times higher than the standard protocol at 120~kV with Filtration~\#2. This increased contrast largely contributes to the enhancement of the detectability index, as discussed in the following subsection.

\begin{figure*}[t!]
\centering
\includegraphics[width=\textwidth]{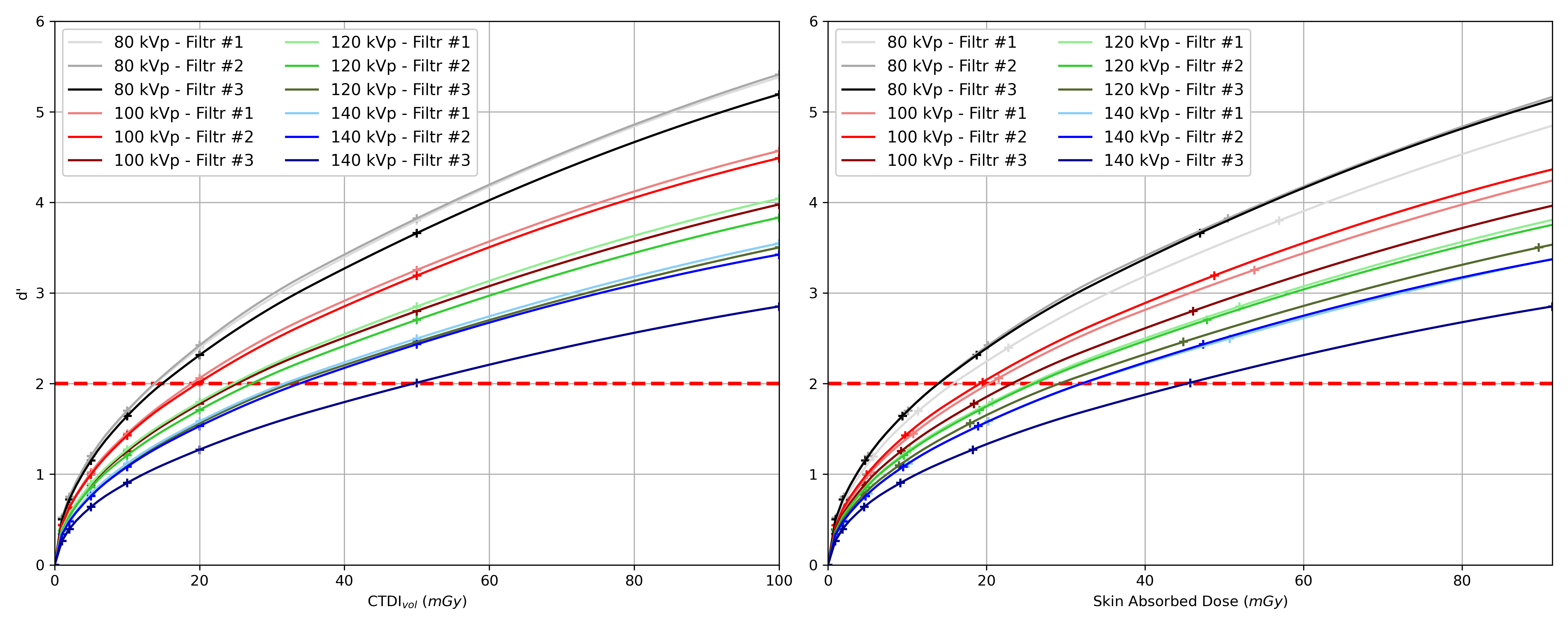}
\caption{Detectability index $d'$ for different CTDI\textsubscript{vol} (Left) and skin absorbed dose (Right). The '+' markers represents the measurement points and the curve are the quadratic interpolation. The red dotted horizontal line figures the level $d'=2$. (For interpretation of the references to colour in this figure legend,  the reader is referred to the web version of this article.)}
\label{fig:d_prime_ctdi_skin}
\end{figure*}

\begin{figure*}[t!]
\centering
\includegraphics[width=\textwidth]{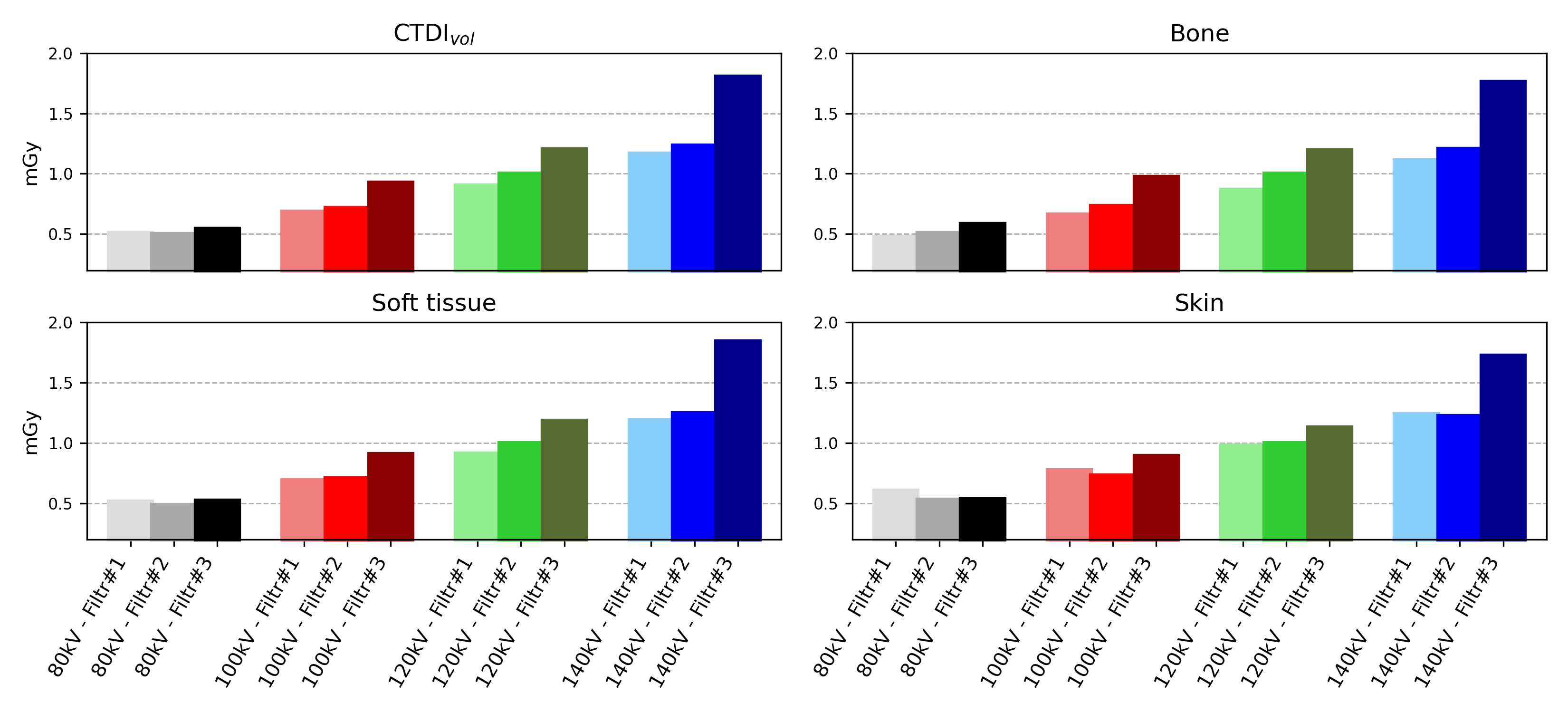}
\caption{CTDI\textsubscript{vol} level (Top Left) and Bone (Top Right), soft tissue (Bottom Left) and skin (Bottom Right) absorbed dose values corresponding to a $d'=2$. For each plot, the values were normalized to the 120~kV-Filtr\#2 case.}
\label{fig:d_prime_organ}
\end{figure*}

\subsubsection{Detectability index}
The detectability index ($d'$) curves are displayed in \hyperref[fig:d_prime_ctdi_skin]{\cref{fig:d_prime_ctdi_skin}} as a function of CTDI\textsubscript{vol} (left) and skin dose (right). On the left graph, the value of $d'$ increases with CTDI\textsubscript{vol}, although the rate of increase diminishes at higher dose levels. On the right graph, the CTDI\textsubscript{vol} values are converted into skin absorbed dose, resulting in a distortion of the x-axis and a modification of the curve shapes. In terms of CTDI\textsubscript{vol}, the 80~kV spectra with Filtration~\#1 or \#2 exhibit the highest $d'$ values. However, due to the increase in skin DCC with low filtration, the 80~kV and Filtration~\#1 case is no longer optimal when plotted against skin dose. A similar trend is observed at 100~kV. Nonetheless, the detectability index increases with decreasing source voltage. 

In \hyperref[fig:d_prime_organ]{\cref{fig:d_prime_organ}}, CTDI\textsubscript{vol} and organ absorbed dose values required to reach a detectability index of $d' = 2$ are presented for all incident spectra. These values are normalized to the standard spectrum (120~kV-Filtr\#2). The required dose is generally lower at 80~kV; however, the behavior varies depending on the selected dose indicator. Overall, CTDI\textsubscript{vol} is correlated with the organ dose, although opposite trends can be observed in some cases. Notably, CTDI\textsubscript{vol} does not capture the difference in skin dose between Filtration~\#1 and~\#2 at 80~kV. Furthermore, CTDI\textsubscript{vol} underestimates the extent to which the bone absorbed dose increases with filtration. In contrast, the variation in soft tissue dose generally mirrors that of CTDI\textsubscript{vol}. Compared to the standard protocol, the required dose to achieve $d' = 2$ can be reduced by half or increased by up to 1.5 times, depending on the energy spectrum.

\vspace{3.5cm}

\section{Discussion}

This work demonstrates that spectral shaping can reduce the dose by operating the source at 80~kV for a forearm protocol, while maintaining comparable image quality. Although the DCC for all organs increases at low voltage, this effect is counterbalanced by the increased contrast at lower energies, which enables improved image quality. A limitation at these energies is BH. Even though BH can be corrected prior to reconstruction, high levels of BH may introduce artifacts. In this study, only conventional imaging was investigated. But PCD-CT can also provide spectral information. This spectral information could be limited  while operating at 80~kV, reducing the spectral image quality. This could lead to a different optimal source parameters. Finally, a lower voltage requires a higher dose current to maintain the CTDI\textsubscript{vol} per exposure, imposing higher constraint on the protocol. 

The source voltage enabling the lowest dose required to reach $d' = 2$ is 80~kV, resulting a dose reduction of almost 50\%. However, the optimal source filtration depends on the organ to be preserved. In our case, bone and skin have different optimal values and exhibit trends that differ from those of CTDI\textsubscript{vol}. Soft tissue shows an intermediate behaviour, more closely aligned with CTDI\textsubscript{vol}. While CTDI\textsubscript{vol} allows for an initial heuristic assessment, it does not enable fine-tuning of the optimal spectral shaping parameters, since the outcome depends on the targeted material. In our case, relying solely on CTDI\textsubscript{vol} would yield a satisfactory trade-off. To refine optimization, epidemiological data would be required to determine organ prioritization. This prioritization should also consider patient-specific factors, such as prior radiation exposure or cancer risk data (e.g., gender, age, ...).

\hyperref[fig:ener_absorb]{\cref{fig:ener_absorb}} displays the mass energy absorption coefficients for the three materials, as well as for air, whose kerma defines the CTDI\textsubscript{vol} value. Soft tissue, skin, and air exhibit similar energy-dependent coefficients, whereas bone has a distinct energy dependence. This highlights that absorbed dose depends not only on material properties but also on anatomical location. In our case, soft tissue and skin share similar energy absorption characteristics but yield different absorbed doses. This discrepancy is observed because skin is located at the forearm's surface, where the photon mean energy is lower. We have demonstrated that skin dose is reduced when decreasing the voltage below 120~kV and increasing the filtration. Similar trends have been demonstrated in recent studies in radiography for newborns~\cite{papadakis2023effect}, chest~\cite{Peglow_2023,aksit2023effect} or abdominal~\cite{Kawashima_2017,aksit2023effect} protocols. The skin is the first material to attenuate incoming photons, particularly those at lower energies. In contrast, bone is situated at the center of the forearm and primarily interacts with photons of higher mean energy. Consequently, bone is less affected by low-energy photons and limited benefit from additional source filtration was observed.

This study presents several limitations, notably in the detector modeling, observer model and CT reconstruction. In our case, images were reconstructed from the photon fluence scored at the detector. This corresponds to an ideal photon-counting detector (PCD) model, which does not account for charge-sharing or fluorescence events. Although such events may degrade image quality, their impact is reduced when no energy thresholds are applied (i.e., when all incoming photons are counted in a single energy bin to reconstruct a conventional image).  

The detectability index $d'$ results depends also on the selected parameters for the observer model. In this work, the model was tuned to correspond to a specific task: the bone delineation in a water-like background with 0.25~mm details. The model could be adapted to other relevant clinical task such as gout diagnosis~\cite{stamp_2019,mourad2024chances}. It has been demonstrated that the optimal spectral shaping parameters can depend on the targeted clinical task~\cite{steidelDoseReductionPotential2022}.

Another limitation lies in the CT reconstruction method. The implemented algorithm is an efficient iterative reconstruction with spatial regularization. The results depend on the selected regularization parameter $\lambda$, which controls the trade-off between image noise and spatial resolution. In this study, the same $\lambda$ value was used across all spectra. However, this parameter could be optimized for each incident spectrum since contrast varies with the mean spectrum energy. The optimal $\lambda$ should ideally be selected in collaboration with radiologists through visual inspection. This topic remains an active area of research~\cite{zhang2018regularization} and is beyond the scope of the present study.

The results also depend on the denoising technique. In this work, denoising was achieved through iterative reconstruction. Recent advances in artificial intelligence (AI)-based methods have shown potential to outperform traditional iterative reconstruction techniques~\cite{jiangDeepLearningReconstruction2022,racineTaskbasedCharacterizationDeep2020,muckSaferImagingComparative2024}. Enhanced denoising methods could further benefit from spectral shaping conditions that lead to higher contrast. Most denoising algorithms exhibit non-linear behavior with respect to noise, resulting in less degradation of spatial resolution under high-noise, high-contrast conditions. In our case, this would likely reinforce the preference for operating the X-ray source at 80~kV, which maximizes the contrast.

The present study aims to optimize the MARS Extremity 5X120 forearm scanner, which features a FOV of 120~mm. In this work, the forearm tissues have been categorized into three classes. However, other anatomical regions contain more radiosensitive organs, such as the thyroid, breasts or lungs. Furthermore, the optimal parameters for spectral shaping are dependent on the WED of the patient, with larger patients benefiting from higher voltages~\cite{steidelDoseReductionPotential2022,suntharalingam2018spectral}. In adult patients, the wrist WED is not expected is not expected to exhibit significant variation between individuals. Full-body scanners, which have a 500~mm FOV, are associated with higher and more varied WED values. Consequently, the results obtained in this study cannot be directly applied to larger scanners. Nevertheless, a similar methodological approach could be employed, potentially including the patient size as a variable.

\section{Conclusion}

This study demonstrates the potential of spectral shaping at 80~kV to reduce radiation dose while maintaining image quality in forearm CT imaging. The findings highlight the limitations of CTDI\textsubscript{vol} as a universal dose indicator, especially when organ-specific absorbed dose variations are considered, and support a shift toward patient-specific imaging strategies. Future work will focus on experimental validation to confirm these simulation-based results with real date measured by a MARS Extremity 5X120 forearm PCD-CT.

\section*{Declaration of competing interest}

The authors declare that they have no known competing financial
interests or personal relationships that could have appeared to influence
the work reported in this paper.

\section*{Acknowledgments}

This work was performed within the framework of the Swiss national science foundation grant No. 212764.  

\bibliographystyle{elsarticle-num-names} 
\bibliography{biblio.bib}


\end{document}